\documentclass[sigconf,screen]{acmart}
\usepackage{verbatim}
\usepackage{longtable}
\usepackage{adjustbox}
\usepackage{multirow}
\usepackage{amsmath,amsfonts}
\usepackage{fancyhdr}
\usepackage[ruled,vlined,linesnumbered]{algorithm2e}
\usepackage{graphicx}
\usepackage{textcomp}
\usepackage{xcolor}
\usepackage{subcaption}
\usepackage{algpseudocode}
\usepackage{enumitem}
\usepackage{float}
\usepackage{graphicx}
\usepackage[normalem]{ulem}
\usepackage{url}
\useunder{\uline}{\ul}{}
\makeatletter
\let\@authorsaddresses\@empty
\makeatother
\AtBeginDocument{%
  }

\begin{document}

\copyrightyear{2025}
\acmYear{2025}
\setcopyright{cc}
\setcctype{by}
\acmConference[FSE Companion '25]{33rd ACM International Conference on the Foundations of Software Engineering}{June 23--28, 2025}{Trondheim, Norway}
\acmBooktitle{33rd ACM International Conference on the Foundations of Software Engineering (FSE Companion '25), June 23--28, 2025, Trondheim, Norway}\acmDOI{10.1145/3696630.3728517}
\acmISBN{979-8-4007-1276-0/2025/06}

\title{\textit{SmartShift}: A Secure and Efficient Approach to Smart Contract Migration}


\author{Tahrim Hossain}
\email{mhossa22@syr.edu}
\affiliation{%
  \institution{Syracuse University}
  \city{Syracuse}
  \state{NY}
  \country{USA}
}

\author{Faisal Haque Bappy}
\email{fbappy@syr.edu}
\affiliation{%
  \institution{Syracuse University}
  \city{Syracuse}
  \state{NY}
  \country{USA}
}

\author{Tarannum Shaila Zaman}
\email{zamant@umbc.edu}
\affiliation{%
  \institution{University of Maryland, Baltimore County}
  \city{Baltimore}
  \state{MD}
  \country{USA}
}

\author{Raiful Hasan}
\email{rhasan7@kent.edu}
\affiliation{%
  \institution{Kent State University}
  \city{Kent}
  \state{OH}
  \country{USA}
}

\author{Tariqul Islam}
\email{mtislam@syr.edu}
\affiliation{%
  \institution{Syracuse University}
  \city{Syracuse}
  \state{NY}
  \country{USA}
}


\renewcommand{\shortauthors}{Hossain et al.}

\begin{abstract}
Blockchain and smart contracts have emerged as revolutionary technologies transforming distributed computing. While platform evolution and smart contracts' inherent immutability necessitate migrations both across and within chains, migrating the vast amounts of critical data in these contracts while maintaining data integrity and minimizing operational disruption presents a significant challenge. To address these challenges, we present \textit{SmartShift}, a framework that enables secure and efficient smart contract migrations through intelligent state partitioning and progressive function activation, preserving operational continuity during transitions. Our comprehensive evaluation demonstrates that \textit{SmartShift} significantly reduces migration downtime while ensuring robust security, establishing a foundation for efficient and secure smart contract migration systems.
\end{abstract}



\maketitle

\section{Introduction}

Blockchain technology has transformed distributed computing through enhanced transparency, security, and decentralization. While initially focused on cryptocurrencies, its impact now spans finance \cite{adams2021uniswap}, healthcare \cite{rifi2017towards}, and digital identity verification \cite{8802771}. Smart contracts \cite{szabo1996smart},self-executing programs on blockchain networks enable automated business logic and have facilitated the development of decentralized applications managing significant digital assets.

The blockchain ecosystem has diversified remarkably over the past decade. While Ethereum \cite{buterin2013ethereum} pioneered programmable smart contracts, newer platforms like Solana \cite{yakovenko2018solana} and Polkadot \cite{wood2016polkadot} have emerged with distinct priorities like transaction processing speed and cross-chain interoperability. This evolution forces development teams to make crucial decisions as platforms that once met their needs may no longer be suitable due to changing requirements and new technological advancements. This evolution introduces a complex challenge of smart contract migration. Blockchain's immutability means that once smart contracts are deployed, they cannot be changed, requiring new versions for any updates or fixes. This immutable nature necessitates another type of migration - deploying and transitioning to new contract versions even when staying on the same blockchain platform.

The scale of these migrations can be staggering. Popular DeFi protocols and NFT marketplaces on Ethereum often manage millions of users and billions in assets, with smart contracts storing critical data. Migrating such contracts demands meticulous attention to data integrity, user experience continuity, and security. These challenges are akin to those in traditional distributed systems, where ensuring minimal disruptions and maintaining operational continuity during process migrations are pivotal to system stability and user trust \cite{douglis1987process, douglis1989experience}. Recent studies on smart contract migration lack practical implementation and evaluation \cite{bandara2020patterns}. Despite some progress with state extraction, challenges like scalability and downtime continue to exist \cite{ayub2023storage, ayub2024sound}. Additionally, current solutions maintain state integrity but use resource-intensive methods that restrict scalability \cite{8751335}. Ongoing security vulnerabilities underscore the critical need for strategies that enhance data integrity and security \cite{tang2023smart}.

To the best of our knowledge, there is no research that thoroughly explores the scale and complexities arising from large-scale smart contract migrations in a comprehensive and actionable way. Understanding the vast scale of such migrations is fundamental to ensuring data integrity, while also developing strategies that effectively ensure security and minimize long downtimes. If migration strategies do not address the massive scale of data and the complexities associated with it, they risk causing substantial disruptions and potential losses. Conversely, if these strategies are not optimized for efficiency, the prolonged downtimes could discourage the broader adoption and trust in smart contract applications.

To address these challenges, this paper introduces \textit{SmartShift}, a framework that optimizes smart contract migrations by intelligently segmenting smart contract data based on its dependencies with key functionalities and the priority of those functionalities to ongoing operations, minimizing disruptions during transitions.

The core contributions of this paper are as follows.
\begin{itemize}
    \item We introduced a protocol for mapping smart contract functions and data relationships, optimizing migration sequences to minimize downtime based on function dependencies and operational importance.

    \item We developed a migration execution mechanism that segments smart contracts into manageable units for efficient blockchain deployment, adhering to gas limits.

    \item We implemented a proof-of-concept\footnote{The prototype of \textit{SmartShift} is available on \url{https://github.com/SmartShift-Anon/SmartShift}} system, evaluated its efficiency across various Ethereum token standards, and conducted a formal security analysis to ensure robust performance and secure migrations.
\end{itemize}

The rest of the paper is organized as follows: Section 2 reviews related literature, Section 3 introduces the design principles of \textit{SmartShift}, Section 4 details its architecture, Section 5 evaluates performance, Section 6 analyzes security, and Section 7 concludes the paper.
\section{Related Work}
Smart contract migrations are vital as the rapid evolution of blockchain platforms necessitates upgrades to newer, more efficient systems. This mirrors the challenges in traditional distributed systems, where seamless process migrations are essential for maintaining system integrity. Similar to research in distributed systems, which focuses on optimizing resource utilization, reducing migration costs, and ensuring swift resumption of processes to minimize downtime \cite{douglis1987process, douglis1989experience}, blockchain migration efforts also strive to facilitate seamless transitions across platforms while maintaining integrity and minimizing operational disruptions. 

Current research in smart contract data migration has advanced by identifying and categorizing migration patterns, providing a theoretical foundation for developers and researchers alike \cite{bandara2020patterns}. These studies have improved the understanding of migration dynamics but frequently lack detailed algorithmic specifications, practical implementations, and thorough performance evaluations for real-world applications. Progress in areas like smart contract state extraction has advanced, though these methods only tackle a fraction of the broader migration challenges \cite{ayub2023storage,ayub2024sound}. Solutions for migrating smart contracts between EVM-compatible platforms have laid the groundwork for maintaining state integrity and deploying efficient migration strategies but face practical limitations due to their reliance on resource-intensive processes which impose scalability constraints and increase gas consumption \cite{8751335}. Further investigation into migrations between Ethereum and newer platforms like Arbitrum has uncovered critical vulnerabilities related to differences in execution environments, including outdated off-chain data retrieval and susceptibility to denial-of-service (DoS) attacks, highlighting the complex security challenges that these migrations present \cite{tang2023smart}. Despite these advancements, significant gaps remain in the development of advanced migration frameworks that effectively address the scalability, security, and complexity of migrating smart contracts on a large scale.
\section{Formal Foundation}
Smart contracts automate and enforce agreements on blockchains, providing secure, transparent, and tamper-proof operations. Defined as \( C = (F, D, A) \), a smart contract comprises functions \( F \), data \( D \), and a unique blockchain address \( A \). Blockchain's immutability necessitates deploying a new contract \( C' = (F', D', A') \) for any modifications, involving data migration \( \phi: D \rightarrow D' \).

Each data element \( d_i \) in \( D \), the total data set to be migrated, requires computational effort \( g(d_i) \) which consumes \textit{gas}, a unit that measures the computational work needed to execute transactions on the blockchain. The total gas \( G_{\text{total}} \) required is given by:
\(
G_{\text{total}} = \sum_{i=1}^{|D|} g(d_i)
\). If \( G_{\text{total}} \) exceeds the gas limit \( G_{\text{limit}} \), the migration must be split into \( \lceil G_{\text{total}}/G_{\text{limit}} \rceil \) transactions. Breaking down smart contract data migration into multiple parts due to gas limits can prolong migration time, leading to extended downtime that disrupts critical business functions and potentially causes revenue loss and reduced trust.

\textit{SmartShift} optimizes migration by breaking down data transfers into manageable chunks and aligning function availability with data readiness, maintaining continuous operation during upgrades. The system uses a dependency matrix $\Delta$ and function priority vector $P$ to orchestrate migrations effectively. $\Delta$ maps function-data dependencies \(\Delta_{f_i \times d_j} = 1\) if function \(f_i\) depends on data \(d_j\); otherwise, \(\Delta_{f_i \times d_j} = 0\).


This matrix is crucial for determining the order of data migration. It clearly identifies which data elements must be migrated and made ready before their dependent functions can be activated, while \(P = [P(f_1), P(f_2), \ldots, P(f_n)]\) ranks function importance based on usage and criticality metrics. Together, these tools determine the optimal migration sequence and ensure data availability aligns with function activation.

\section{System Architecture}
\begin{figure*} [t]
    \centering
    \includegraphics[width=0.75\textwidth]
{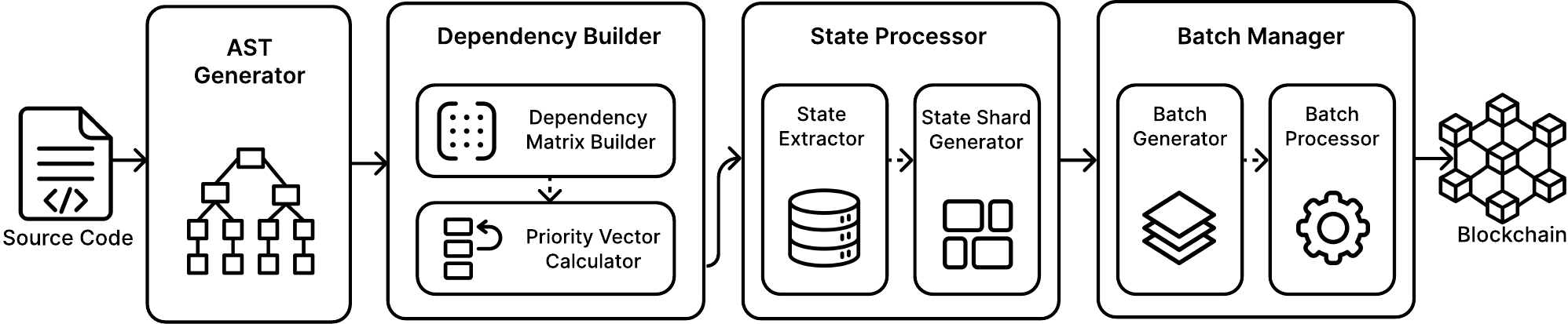}
    \caption{System Architecture} 
    \label{fig:workflow} 
\end{figure*}

The \textit{SmartShift} architecture consists of interconnected components that analyze smart contract dependencies and coordinate segmented migrations to ensure continuity. This section explains their interaction within the migration workflow (Fig. \ref{fig:workflow}).

\subsection{AST Generator}
We have implemented an AST Generator to parse the smart contract's source code into an Abstract Syntax Tree (AST). An AST organizes the code into a tree structure, where each node represents a different programming construct. The hierarchical tree structure of the AST simplifies analysis by clearly organizing the smart contract's code, making it easier to navigate and break down complex elements. We then leverage this structured representation in our Dependency Builder to efficiently organize and prioritize migration tasks.
\subsection{Dependency Builder}
The Dependency Builder analyzes and maps the dependencies between functions and data elements within the smart contract. It effectively prioritizes these elements based on their interdependencies and operational importance, ensuring that migrations are executed in a sequence that minimizes downtime and maintains functionality. This module is comprised of two integral components, which are discussed in further detail below.
\subsubsection{Dependency Matrix Builder}
We have implemented the Dependency Matrix Builder to construct a matrix to map the dependencies between functions and state variables within a smart contract. This precise mapping allows us to effectively orchestrate the migration process by ensuring that functions can be activated immediately after their dependent data elements are successfully migrated. This capability significantly reduces downtime by maintaining operational continuity, as functions resume operation as soon as the necessary data elements are in place, thus enhancing the overall efficiency and integrity of the migration process.
\subsubsection{Priority Vector Calculator}
We have implemented a Priority Vector Calculator to strategically determine the migration order of smart contract functions based on their usage frequency and operational significance. Our process involves collecting data on function call frequency and criticality, assigning importance-based weights, and aggregating these into priority scores. We then rank functions using these scores to create a priority vector that dictates the optimal migration sequence. This method ensures critical functions are migrated first, minimizing disruptions and maintaining operational continuity.
\subsection{State Processor}
The State Processor in the \textit{SmartShift} framework analyzes smart contract storage and segments it into manageable units for efficient migration. It consists of two key modules: the State Extractor and the State Shard Generator. These modules are discussed below.
\subsubsection{State Extractor}
We have designed the State Extractor to analyze smart contract source code, enabling it to accurately map the storage structure of Ethereum smart contracts. In Ethereum, smart contracts store state data in a structured format known as the Storage Trie, which consists of $2^{256}$ 32-byte slots. These slots are allocated based on the declaration order in the source code and size of state variables. Upon analyzing the smart contract source code, the State Extractor outputs detailed instructions about where and how each data element is stored within these slots. Specifically, it notes the type of each variable, its precise location in the storage trie (slot number), and its offset within that slot if the data does not occupy the entire slot. These instructions are pivotal for the State Shard Generator, that uses them to segment the smart contract state into smaller, manageable units.
\subsubsection{State Shard Generator}
The State Shard Generator precisely segments the smart contract’s state into individual shards, leveraging the detailed instructions provided by the State Extractor. Each shard encapsulates a single state variable or data element in the smart contract. This process involves selecting the appropriate slots from the contract's storage, as defined by the State Extractor's output, and partitioning them into discrete units or shards. Each shard is thus a self-contained packet of slots related to a state variable. This approach allows for targeted migrations where specific portions of the contract’s state can be updated or moved independently of others.

\textbf{Batch Manager.} The Batch Manager is a critical component of \textit{SmartShift}, responsible for organizing and executing the migration process in a structured and efficient manner. As illustrated in Fig. \ref{fig:workflow} it consists of two subcomponents: the Batch Generator and the Batch Processor.

\textbf{Batch Generator.} We have developed the Batch Generator to sort the shards generated by the State Shard Generator using the dependency matrix and the priority vector. This ensures that interdependencies and operational importance are respected throughout the process. It adjusts the composition of shards as needed, either merging multiple shards together or breaking them down further. These adjustments are guided by gas limit constraints and user preferences. This flexibility allows us to create optimally sized batches that balance performance and resource usage while maintaining the integrity of the migration process.

\textbf{Batch Processor.} We encapsulate each batch into a blockchain transaction and sequentially send them to the blockchain using the Batch Processor.  We have implemented It to process the batches in the order determined by their priority, ensuring that critical functions and data are migrated first. By encapsulating batches into transactions, the Batch Processor ensures that each migration step is executed securely and efficiently.

\section{Implementation \& Evaluation}
To measure the performance of the solution, we have implemented a proof of concept for \textit{SmartShift}. We implemented \textit{SmartShift}'s key components in a combination of Python and Go. To access transaction history and gas limit required to implement the Priority Vector Calculator and the Batch Processor, we have used services like Infura and Etherscan. Using this proof of concept, we analyzed how effectively our framework reduces migration-related downtime. For quantitative evaluation, we developed a metric called the Function Activation Threshold (FAT), which measures the average number of data elements required to activate smart contract functions. Lower thresholds indicate quicker migrations, while higher ones suggest longer unavailability. We analyzed this metric across the ERC-20, ERC-721, and ERC-1155 Ethereum token standards, assessing their impact on operations due to their critical role and widespread use in decentralized applications. 

\begin{figure}[h]
\centering
\includegraphics[width=0.68\columnwidth]{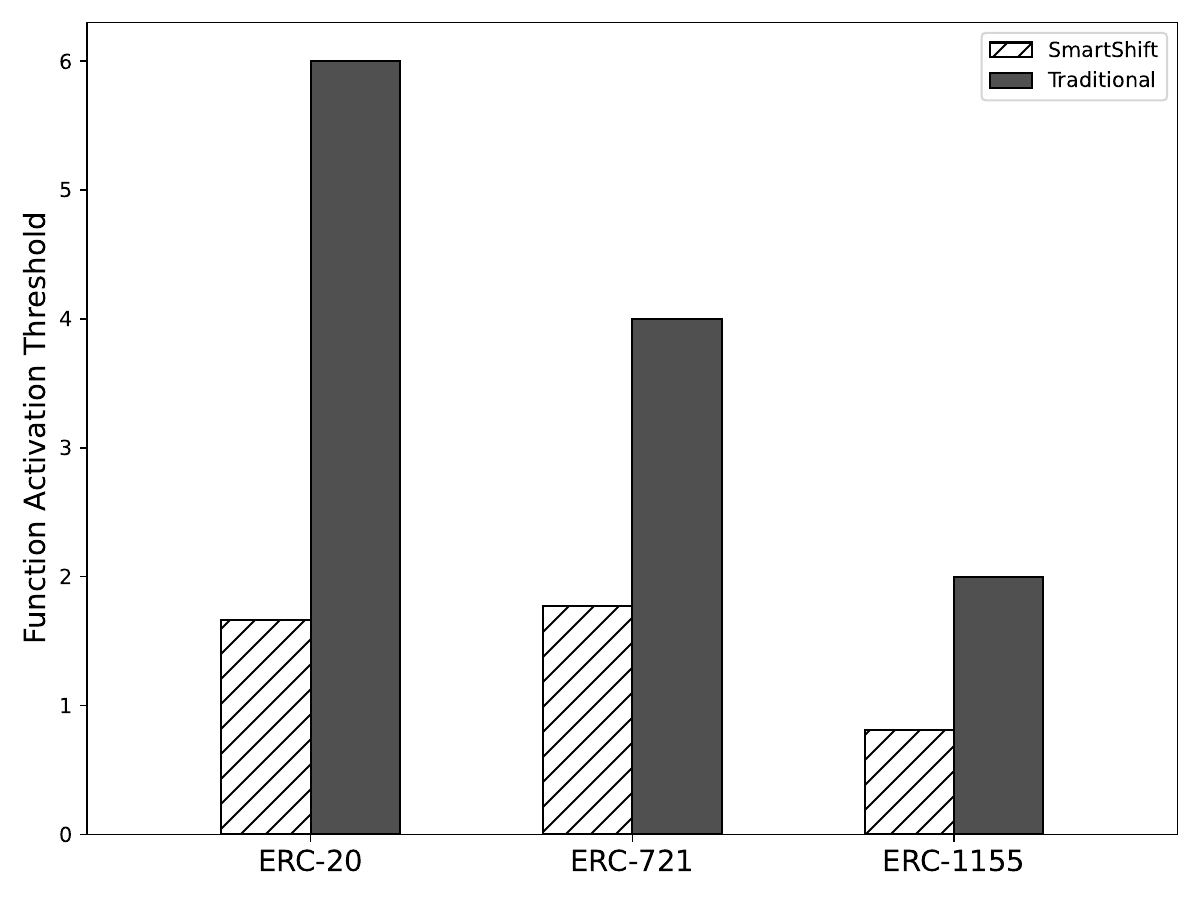}
\caption{Comparison of Function Activation Threshold}
\label{fig:fatcomparison}
\end{figure}

We used the FAT metric to compare \textit{SmartShift}'s performance with traditional migration methods that rely on constructor-based initialization or monolithic functions for state initialization after migration. For \textit{SmartShift}, we calculated FAT using our dependency matrix, which precisely maps each function to its required data elements, allowing us to compute the average number of data dependencies per function. Traditional methods result in higher thresholds due to their single atomic operations during deployment. \textit{SmartShift}'s incremental approach achieves lower thresholds across all tested standards, as shown in Fig. \ref{fig:fatcomparison}, reducing FAT by 72\% for ERC-20, 55\% for ERC-721, and 60\% for ERC-1155 tokens.

\section{Security Analysis}
Security is a central focus of \textit{SmartShift}, which meticulously addresses the multifaceted risks inherent in smart contract migrations through robust and proactive measures. Our approach leverages automated dependency mapping and priority-based segmented migration processes to ensure comprehensive security throughout the migration lifecycle. These features collectively establish \textit{SmartShift} as a secure and resilient framework for smart contract upgradability.

In migration scenarios, non-optimized transitions can act like a Denial of Service (DoS) attack, temporarily disabling critical functions and disrupting operations across dependent contracts. This can cause extensive system-wide dysfunction, particularly in environments with complex and interdependent functions where one contract's failure triggers a domino effect, halting or impairing subsequent operations across multiple contracts. \textit{SmartShift} addresses this vulnerability through its intelligent data dependency and segmentation procedures, which analyzes function dependencies and operational priorities to create optimized migration pathways.

\textit{\textbf{Proposition 1}}: Unoptimized migrations misprioritize critical dependencies \( \Delta_{f_{\text{critical}} \times d_{\text{critical}}} = 1 \), leading to extended downtimes \( T_{\text{downtime}}(f_{\text{critical}}) > T_{\text{acceptable}} \), exposing systems to DoS-like vulnerabilities. \textit{SmartShift} mitigates this by ensuring \( T_{\text{delays}}(d_{\text{critical}}) \approx 0 \) through structured prioritization, preserving system reliability and integrity.

\textit{\textbf{Analysis}}: In scenarios lacking an optimized migration strategy, critical interdependencies \( \Delta \) and priorities \( P \) are neglected, leading to the overlooked prioritization of essential dependencies \( \Delta_{f_{\text{critical}} \times d_{\text{critical}}} = 1 \) in favor of non-essential ones \( \Delta_{f_{\text{critical}} \times d_{\text{noncritical}}} = 1 \). This misalignment results in the prioritization of non-critical data \( d_{\text{noncritical}} \) over critical data \( d_{\text{critical}} \), causing extended downtime:
\[
T_{\text{downtime}}(f_{\text{critical}}) = T_{\text{migration}}(d_{\text{critical}}) + T_{\text{delays}}(d_{\text{critical}}),
\]
where \( T_{\text{delays}} \) is the delay caused by incorrect data prioritization. If \( P(f_{\text{critical}}) > P(f_{\text{noncritical}}) \) is not maintained, then \( T_{\text{downtime}}(f_{\text{critical}}) \) exceeds \( T_{\text{acceptable}} \), enhancing the risk of DoS-like disruptions.

\textit{SmartShift} systematically aligns migration activities with the dependencies \( \Delta_{f_i \times d_i} \) and priorities \( P(f_i) \) of the system functions. By ensuring \( \Delta_{f_{\text{critical}} \times d_{\text{critical}}} = 1 \) and that \( P(f_{\text{critical}}) > P(f_{\text{noncritical}}) \), it guarantees that critical functions \( f_{\text{critical}} \) resume operations promptly post-migration, thus effectively minimizing the downtime:
\[
T_{\text{downtime}}(f_{\text{critical}}) = T_{\text{migration}}(d_{\text{critical}}),
\]
with minimal delays \( T_{\text{delays}}(d_{\text{critical}}) \approx 0 \). This prioritization ensures that \( T_{\text{downtime}}(f_{\text{critical}}) \) remains within acceptable limits, thereby mitigating prolonged disruptions and vulnerabilities during migration processes.

Improper migration of a smart contract's state can lead to data loss and corruption, undermining the integrity and functionality of the contract. \textit{SmartShift}'s State Processor addresses this issue by ensuring accurate and complete data transfer and reconstruction, thus maintaining the contract’s operational continuity and security.

\textit{\textbf{Proposition 2}}: The \textit{SmartShift} State Processor ensures data integrity by constructing shards \( G_s \) that fully cover all storage slots for each variable, guaranteeing the reconstructed state map \( S' \), which represents the migrated state, satisfies \( S' = S_{\text{original}} \).

\textit{\textbf{Analysis}}: Smart contract storage operates as a key-value store, where each slot \( S_i = \{ K_i, V_i \} \) is identified by a unique key \( K_i \) and stores a value \( V_i \). Slots are allocated to variables based on declaration order and type-specific rules.

In \textit{SmartShift}, shards \( G_s \) are constructed to group all slots related to a single state variable:
\(
G_s = \{ S_i \mid S_i \mapsto v \}.
\)
Improperly constructed shards that miss slots can result in incomplete migrations, leading to undefined behavior, data corruption, or logical errors.

The \textit{SmartShift} State Processor ensures integrity by representing each data element as:
\(
D_i = \{ \text{label}_i, \text{type}_i, \text{slot}_i, \text{offset}_i \},
\)
and categorizing data types as:
\(
T = \{ \text{type}, \text{base}, \text{encoding}, \text{number of bytes} \}.
\)

Using this information, shards are generated to fully cover all slots for each variable, ensuring:
\(
S' = \bigcup_{s} G_s.
\)
The final step validates the new state map
\(
S' = S_{\text{original}},
\)
ensuring no data is lost or misaligned.

\section{Conclusion}
In this paper, we introduced \textit{SmartShift}, a sophisticated framework that addresses key challenges related to smart contract migration security, efficiency, and scalability by employing granular state decomposition and incremental function activation. Our empirical tests demonstrate that \textit{SmartShift} significantly reduces system downtime and strengthens defenses against common migration vulnerabilities. Through these innovations, \textit{SmartShift} defines the next step in blockchain development, focusing on secure, seamless transitions for smart contracts.

\bibliographystyle{ACM-Reference-Format}
\bibliography{sample-base}


\end{document}